\documentclass[]{spie}

\usepackage[]{graphicx}

\usepackage{dcolumn}
\usepackage{bm}

\newcommand{\ket}[1]{\ensuremath{\left|#1\right\rangle}}

\title{A Compact Source for Quantum Image Processing with Four-wave Mixing in Rubidium-85}

\author{Ulrich Vogl, Ryan Glasser and Paul D. Lett
\skiplinehalf
National Institute of Standards and Technology and\\
Joint Quantum Institute, NIST \& the University of Maryland,\\ Gaithersburg, MD 20899 USA\\
}

\authorinfo{ E-mail: ulrich.vogl@nist.gov}

\begin{document}
  \maketitle

\begin{abstract}
We have built a compact light source for bright squeezed twin-beams at 795\,nm based on four-wave-mixing in atomic $^{85}$Rb vapor. With a total optical power of 400\,mW derived from a free running diode laser and a tapered amplifier to pump the four-wave-mixing process, we achieve 2.1\,dB intensity difference squeezing of the twin beams below the standard quantum limit, without accounting for losses.  Squeezed twin beams generated by the type of source presented here could be used as reference for the precise calibration of photodetectors.  Transferring the quantum correlations from the light to atoms in order to generate correlated atom beams is another interesting prospect. In this work we investigate the dispersion that is generated by the employed four-wave-mixing process with respect to bandwidth and dependence on probe detuning.  We are currently using this squeezed light source to test the transfer of spatial information and quantum correlations through media of anomalous dispersion.

\end{abstract}


\keywords{Squeezed light, Four-wave-mixing, Anomalous dispersion}

\section{INTRODUCTION}
\label{sec:intro}  

In recent years  quantum correlations
and entanglement present in multi-mode squeezed fields have generated much interest, as such states can have interesting
applications in quantum information, quantum imaging, and quantum
computing \cite{slusher1985, 1994OptCo.104..374K, 2002PhRvA..65a3813N, 2001PhRvL..86.4267S, 2004PhRvA..70d2315L, 2005RvMP...77..513B, 2005PhRvA..72a3802H, 2005PhRvL..94v3603M, rolston2010,2010Grangier}.\\
We use a non-degenerate four-wave mixing (4WM) process in
rubidium-85 vapor at 795 nm to generate correlated and quantum-mechanically entangled spatially multi-mode twin beams \cite{lett2007, lett2008, alberto2008, 2011Neil}.
Earlier work based on titanium-sapphire laser systems as the primary light source has shown the generic advances of this particular source of multi-mode squeezed light.  High levels of squeezing competitive with sources based on parametric down-conversion have been shown, without the need for a build-up cavity.  We have built a compact 4WM source of squeezed light using only semiconductor lasers.

\section{EXPERIMENTAL SCHEME AND SETUP}
We employ the double-lambda scheme in atomic rubidium vapor shown in Fig.\,1 that has been described in greater detail elsewhere \cite{lett2007}.
 A strong, linearly polarized laser beam acts as pump and is detuned $\Delta\approx$800\,MHz to the blue of the Rb D1 line at $\approx795$\,nm, $\ket{5S_{1/2},F=2}\rightarrow\ket{5P_{1/2}}$, as shown in Fig.\,1.  A weak, orthogonally polarized probe is detuned $\approx3$\,GHz to the blue of the pump beam and injected at an angle of $\approx1^{\circ}$ relative to the pump.  A conjugate pulse is created symmetric to the probe on the other side of the pump, in accordance with the relevant phase-matching conditions of the process.

 The experimental implementation is shown in Fig.\,1. The pump and probe beams are initially derived from a tunable diode laser with a maximum output power of 110\,mW. Approximately half this output is fed single pass into a tapered amplifier chip and amplified to $\approx 400$\,mW. The output of the tapered amplifier is subsequently spatially filtered to ensure a more uniform spatial pump profile. The probe beam is double-passed through a 1.5\,GHz acousto-optic modulator, which detunes the probe relative to the pump by roughly the 3\,GHz ground state splitting.
   \begin{figure}
   \begin{center}
   \begin{tabular}{c}
   \includegraphics[width=16.5cm]{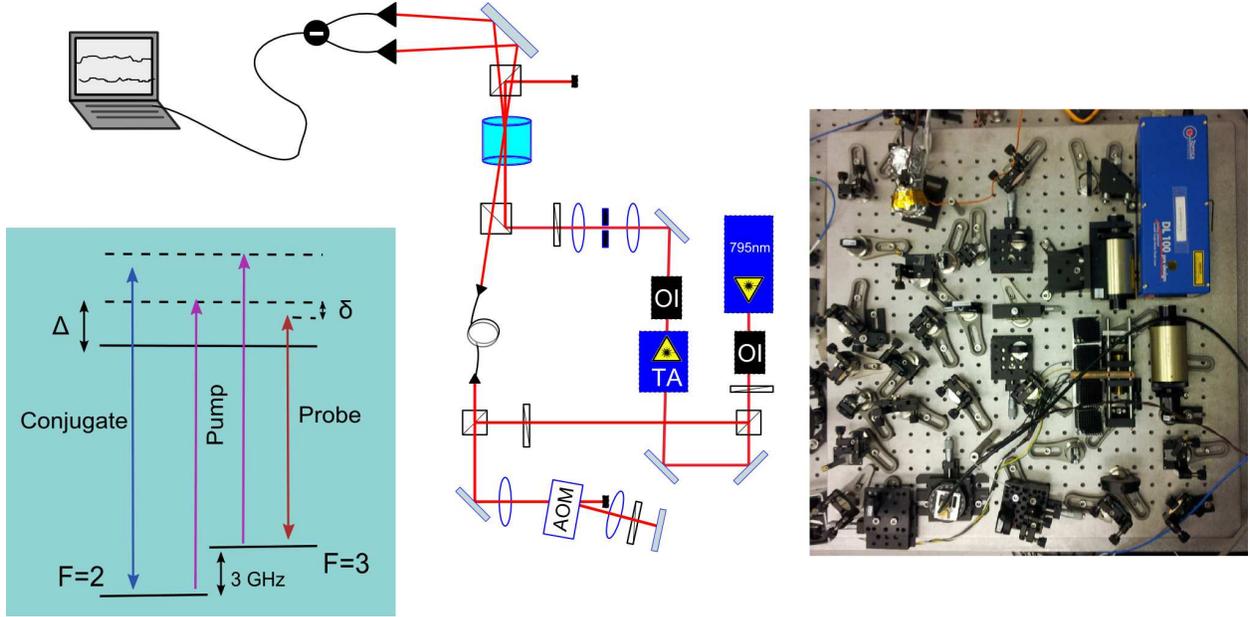}
   \end{tabular}
   \end{center}
   \caption[example]
   { \label{fig:setup}
 (a) Level scheme for the double-lambda system with the relevant one-photon detuning $\Delta$ and the two-photon detuning $\delta$ as indicated. (b) Setup of the compact source. (TA: tapered amplifier, OI: optical isolator, AOM: acousto-optic modulator.)}
   \end{figure}
Both beams are combined on a polarizing beam splitter and sent into the 1.7\,cm long rubidium cell. The cell is anti-reflection coated for 800\,nm and heated to $\approx$115\,$^{\circ}$C, which corresponds to a rubidium atom number density of $\approx 2\cdot 10^{13}$\,cm$^{-3}$.
The pump beam has a 450\,$\mu$m waist, while the probe has a slightly smaller 350\,$\mu$m waist (1/$e^2$ radius). The seeded 4WM process converts two photons from the pump into one photon at the probe frequency and in the probe mode and one conjugate photon at the frequency $\omega_c=2\omega_{\mathrm{pump}}-\omega_{\mathrm{probe}}$. Pump and probe are aligned to intersect with an azimuthal angle $\Theta$ approximately fulfilling the phase-matching condition for the 4WM process; the conjugate photons are accordingly created in a mode with an azimuthal angle $-\Theta$ relative to the pump.

\section{INTENSITY DIFFERENCE SQUEEZING}
The used 4WM process can be described as a phase-insensitive amplifier in which the probe beam is amplified and the conjugate beam intensity is strongly correlated to the intensity of the probe beam. The strong correlation between the twin beams has been demonstrated resulting in a relative intensity squeezing down to 9\,dB below the standard quantum limit \cite{2008McC}. All these results have been obtained with a ring-cavity titanium-sapphire laser source of $\approx$1\,W output power.
In our setup we had the motivation to realize this 4WM process with a diode laser setup, as it allows us to make the whole setup more compact and easier to transport and also less expensive than a titanium-sapphire laser.  Compactness might be useful for experiments where the correlated twin-beams are sent over long distances, and it allows the flexible use of the twin-beam source in multiple experiments.
One possible application is also the use for precise detector calibration based on the strong intensity correlation of the twin beams \cite{1995MetroMigdall, 2011JMOMarino}.
   \begin{figure}
   \begin{center}
   \begin{tabular}{c}
   \includegraphics[width=16.5cm]{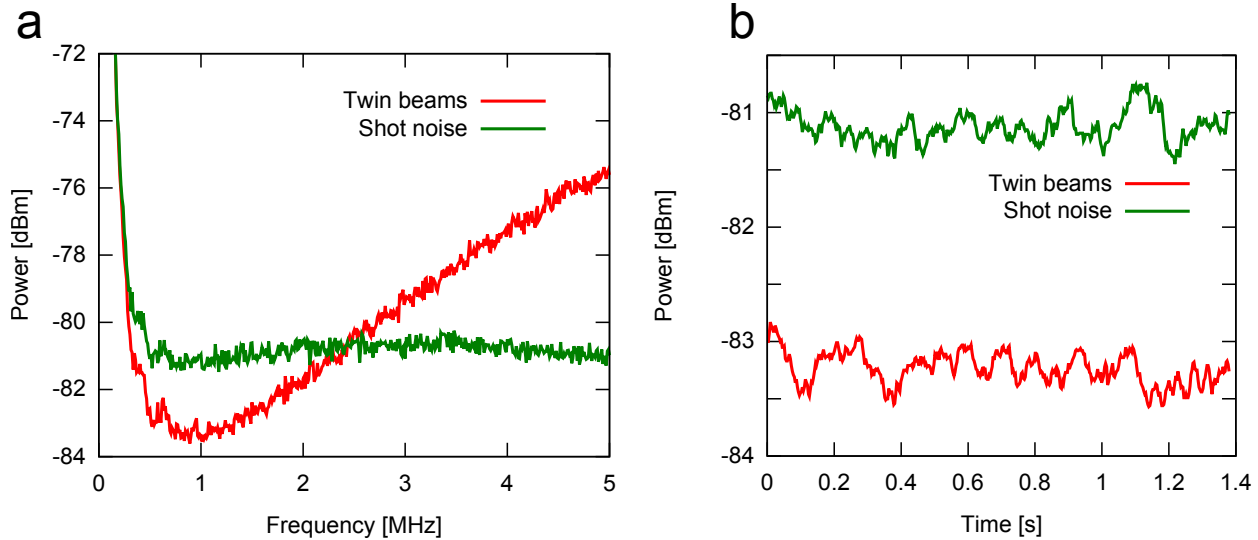}
   \end{tabular}
   \end{center}
   \caption[example]
   { \label{fig:span}
(a) Noise power spectra versus frequency for the experimental shot noise (green) and for the relative intensity noise of the probe and conjugate beam (red) (resolution bandwidth 30\,kHz, video bandwidth 300\,Hz). (b) Zero-span trace for the shot noise and the relative intensity noise at 800\,kHz. An intensity difference squeezing of 2.1\,dB is obtained.}
   \end{figure}

To characterize the relative intensity squeezing that is obtainable in our system we use direct detection with a pair of balanced photodiodes and monitor the difference of the generated photocurrent with a spectrum analyzer.
We can typically achieve a probe gain of 4-10 for an applied probe power of 5\,$\mu$W.
Some of the results obtained are shown in Fig.\,2. Figure 2a shows noise power versus the noise frequency with the reference trace for the shot-noise (green), which marks the standard quantum limit, and the trace for the intensity difference noise of the twin-beams (red). For the region between 200\,kHz and 2\,MHz we observe significant squeezing. The best noise reduction we achieve is around 800\,MHz, for which we show a zero-span trace in Fig.\,2b. The average noise reduction compared to the shot-noise limit is 2.1\,dB, where we did not account for any losses from imperfect optical elements or detection after the cell where the twin beams are generated. The bandwidth for which we can observe squeezing is in agreement with previous experiments performed with ring-cavity laser sources. There it has been shown that the bandwidth where one can obtain squeezing is critically affected by the differential group delay that the twin beams experience. Our system has not been optimized for this, leading to the excess noise above about 2.5\,MHz in Fig.\,2a. The total amount of squeezing is limited by the available pump beam power. The tapered amplifier chip we used to generate the pump beam introduces only a small amount of extra noise, and tapered amplifier chips with higher output power are commercially available.

\section{DISPERSION AND GROUP INDEX FOR THE TWIN BEAMS}
To obtain optimum relative intensity squeezing in the 4WM process, one is restricted (depending on the rubidium number density and other experimental parameters) to a range of the one-photon detuning of $\approx$ 200\,MHz and for the two-photon detuning of a few MHz. The main reasons for this are the differential gain that the probe and the conjugate experience, as one of the beams is close to an absorption line, and, via the different dispersion, the different group velocities that the two beams have in the cell.  This can induce a significant fractional delay between the two beams, which, if not accounted for by an extra delay line before the detection, can obscure the relative intensity squeezing in the measurement as seen in Fig\,2.a.

The varying differential absorption that affects the probe and the conjugate can be seen in Fig.\,3a, where we show transmission line scans for a wide range of one-photon detunings of the pump beam while the probe beam frequency is scanned over a range of 14\,GHz over the relevant structures of the employed rubidium transition. The relative amplitude for the transmission is indicated for the lower trace. The amplified probe is visible as a peak at a positive detuning in the range from 3\,GHz to 4\,GHz. The pump is set in a range $\approx$ 100\,MHz to 1\,GHz blue detuned from the $\ket{5S_{1/2},F=2}\rightarrow\ket{5P_{1/2}}$ transition. While the probe beam in these scans is injected far off resonance and subsequently experiences only gain, the generated conjugate is created between the two strong absorption lines caused by the hyperfine splitting of the ground state.
This configuration of coupled generation of twin beams has been investigated for the situation of slow light propagation for both beams experimentally \cite{boyer2007} and theoretically \cite{Lukin2000b}, which showed that the group velocity of input pulses is locked to each other, where the locking can be varied via detuning and optical density.
   \begin{figure}
   \begin{center}
   \begin{tabular}{c}
   \includegraphics[height=8.5cm]{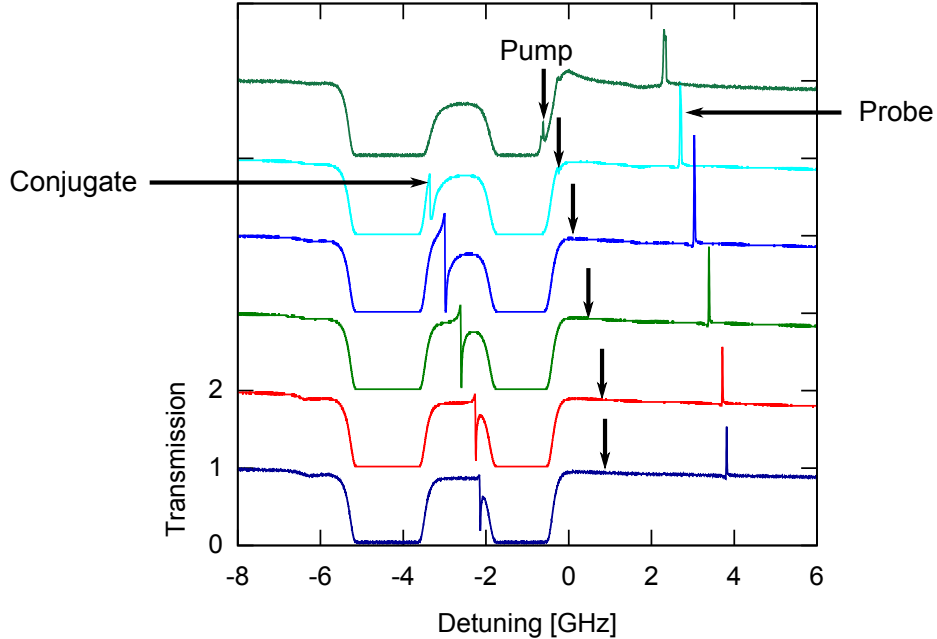}
   \end{tabular}
   \end{center}
   \caption[example]
   { \label{fig:800}
Full line scans for different pump detunings. The relative transmission is indicated for the bottom trace. The detuning is given relative to the optimum pump-detuning for squeezing and the pump frequency is indicated with a vertical arrow for each trace. }
   \end{figure}
The situations where one or both of the beams traverse a region of anomalous dispersion, where the group index of the medium becomes negative, have not been investigated previously.

The complex refractive index of a single gain or absorption line is given by
       \begin{equation}
n(\omega)=n_0+\frac{c \alpha_0}{2 \omega}\frac{\gamma}{\omega-\omega_0+i \gamma},
\end{equation}
where $\alpha_0$ is the absorption or gain coefficient, $\omega_0$ the resonant frequency and $\gamma$ the line width.
The group velocity is accordingly defined by
\begin{equation}
v_g=\frac{c}{n_0+\omega dn/d\omega},
\end{equation}
where $n_0$ is usually close to unity.
In Fig.\,4 we show a close-up of the gain features for the generated (Fig.\,4a) and the injected beam (Fig.\,4b) and the derived group index, $c/v_g$, for a one-photon detuning of 400\,MHz.
Both line shapes can be understood by the model for a coupled double-lambda system given in the work of Lukin et al. \cite{Lukin2000b}. The derived group index gives an estimate for the possible slow- and fast-light effects. The typical frequency bandwidth of the regions of large positive or negative group indices is 5\,MHz - 20\,MHz. For probing these features with optical pulses, the optimum pulse width is in the region of 50\,ns - 200\,ns. Sending in shorter probe pulses with a larger frequency bandwidth is expected to yield a group delay or group advancement that deviates from the group index profiles shown in Fig.\,4, as the pulses sample regions of both normal and anomalous dispersion simultaneously. This may explain why in earlier experiments only a group delay was observed.
The absolute values of the group index shown in Fig.\,4 make this system promising to achieve large relative advancements for both gain features. Further it will be interesting to investigate the coupled propagation of the twin beams under the condition of anomalous dispersion \cite{garrett1970, milonni2002, 2010review}.
   \begin{figure}
   \begin{center}
   \begin{tabular}{c}
   \includegraphics[width=14cm]{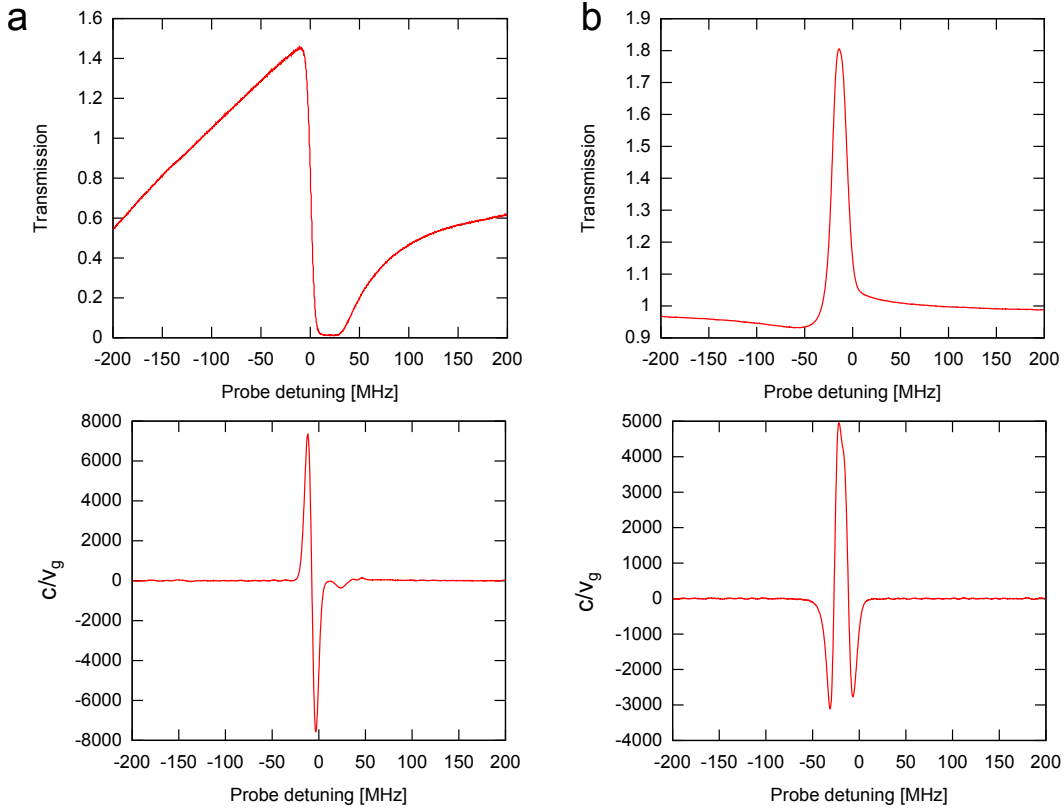}
   \end{tabular}
   \end{center}
   \caption[example]
   { \label{fig:800}
Detailed view of the lineshape of the gain region of the generated beam (a) and the injected beam (b). The detuning is given relative to the respective line centers. In the lower panels we show the group index derived from the experimental line shapes for the generated conjugate and the amplified probe beam. Near resonance of the gain features the group index is strongly modified, while off-resonant $c/v_g$ approaches 1. }
   \end{figure}
Another interesting aspect is that high anomalous dispersion can be achieved with a gain of $\approx$2 and moderate absorption on the generated conjugate line. This is a prerequisite to realize relative intensity squeezing between the twin beams, and can allow investigation into the evolution of the correlated beams when propagating through anomalous dispersion.

\section{SUMMARY}
We have presented a simple and compact setup for the generation of correlated twin beams. A relative intensity squeezing of -2.1\,dB is observed, and scaling of the system is easily obtainable.
Possible future directions in this experiment could be the use of the compact squeezed light source to interact with a Bose-Einstein condensate to produce correlated atom beams \cite{2004JMOLett}.

We also identify and evaluate regions of strong anomalous dispersion created by the 4WM process both for the injected and generated beam. This allows us to consider the interesting possibility of performing  experiments to study the entanglement between the twin beams when one or both of them are passed through a fast-light medium.

\acknowledgments     
This work was supported by the Air Force Office of
Scientific Research. Ulrich Vogl would like to
thank the Alexander von Humboldt Foundation. This research was performed while
Ryan Glasser held a National Research Council Research
Associateship Award at NIST.



\end{document}